\documentclass[sigconf,nonacm]{acmart}
\setcopyright{rightsretained}
\acmConference[IMC '25]{ACM Internet Measurement Conference}{October, 2025}{Location}
\acmDOI{https://doi.org/XX.XXXX/XXXXXXX.XXXXXXX}
\acmISBN{978-1-XXXX-XXXX-X/XX/XX}

\usepackage{amsmath,amssymb,amsfonts}
\usepackage{graphicx}
\usepackage{subcaption}
\usepackage{textcomp}
\usepackage{xcolor}
\usepackage{float}

\usepackage{amsmath} 
\usepackage{amssymb} 

\usepackage[linesnumbered,ruled,vlined]{algorithm2e}

\usepackage{lettrine}
\usepackage{verbatim}
\usepackage{MnSymbol}
\usepackage{tabularx}
\usepackage{soul}
\usepackage{tikz}

\usetikzlibrary{decorations.pathreplacing, positioning, arrows.meta, shapes.multipart}
\usetikzlibrary{shapes,arrows,positioning}
\usepackage[colorinlistoftodos]{todonotes}
\renewcommand\footnotetextcopyrightpermission[1]{} 
\begin{document}

\title{QFAM: Mitigating QUIC Handshake Flooding Attacks Through Crypto Challenges}

\author{Abdollah Jabbari}
\affiliation{%
  \institution{Concordia Institute for Information Systems Engineering (CIISE), \\ Concordia University}
  \city{Montreal}
  \country{Canada}
}
\email{abdollah.jabbari@concordia.ca}

\author{Y A Joarder}
\affiliation{%
  \institution{Concordia Institute for Information Systems Engineering (CIISE), \\ Concordia University}
  \city{Montreal}
  \country{Canada}
}
\email{ya.joarder@concordia.ca}

\author{Benjamin Teyssier}
\affiliation{%
  \institution{Concordia Institute for Information Systems Engineering (CIISE), \\ Concordia University}
  \city{Montreal}
  \country{Canada}
}
\email{benjaminteyssier102@gmail.com}

\author{Carol Fung}
\affiliation{%
  \institution{Concordia Institute for Information Systems Engineering (CIISE), \\ Concordia University}
  \city{Montreal}
  \country{Canada}
}
\email{carol.fung@concordia.ca}

\begin{abstract}
QUIC protocol is primarily designed to optimize web performance and security. However, previous research has pointed out that it is vulnerable to handshake flooding attacks. Attackers can send excessive volume of handshaking requests to exhaust the CPU resource of the server, through utilizing the large CPU amplification factor occurred during the handshake process under attack. 

In this paper, we introduce a novel
defense mechanism by introducing the concept of crypto challenges into the handshake protocol. This enhancement involves a proposal of modifying the RETRY token to integrate a cryptographic challenge into it. The client must solve crypto challenges during the handshake process in order to receive a high priority on the server side. By properly choosing the difficulty level of the challenges,
the CPU amplification can be reduced, thus the DDoS vulnerability is naturalized.

 We evaluated the effectiveness of our proposed solution by integrating the crypto challenges into the clients and server of \textit{aioquic}. Our experimental results demonstrate that our solution can effectively balance the resource usage between the attacker and the server during of handshake flooding attacks while maintaining a low overhead for legitimate clients.
\end{abstract}


\maketitle

\section{Introduction}

QUIC, a next-generation transport protocol, has revolutionized web performance through its enhanced speed, security, and efficiency \cite{mucke2024reacked, mishra2023containing,lorimer2022poster}. QUIC was initially developed as a project by Google in 2012 named "Quick UDP Internet Connections" \cite{roskind2015quic,huang2024quic}. Its purpose was to enhance the web communication's speed, security and efficiency by leveraging UDP for transport \cite{joarder2024exploring, chatzoglou2023revisiting}. Due to its advantages such as reduced latency, improved reliability and security, QUIC is increasingly adopted for multimedia traffic, such as real-time video distributions, on-demand distribution of videos, and interactive applications~\cite{kalan2024survey, barakabitze2019qoe}. Large companies (e.g., Google, Meta, Cloudflare, and Uber) use QUIC to improve the quality and reliability of web services\cite{dey2022detail, wiki_quic_2024}. By 2014, QUIC has been widely deployed in Chrome, becoming a critical component of modern web traffic \cite{chatzoglou2023revisiting}.  According to the recent statistics \cite{quic_usage}, around 8.5\% websites around the world use QUIC as their transport layer protocol as of November 2024. 
QUIC protocol was first standardized in 2021 as RFC 9000~\cite{rfc9000} and it became the transport layer of HTTP/3 \cite{ruth2019perceiving,kuhlewind2021evaluation}. Its second version was released in 2023 as RFC 9369~\cite{rfc9002}, which introduced enhanced migration and compatibility features~\cite{duke_quic_2023}. 

QUIC is mainly designed to reduce latency and improve connection performance \cite{joarder2022survey, teyssier2023empirical}. It performs exceptionally well in environments where traditional protocols such TCP struggle with congestion and long connection setups \cite{chatzoglou2023revisiting}. By integrating Transport Layer Security (TLS) directly into the transport layer, QUIC enables faster handshakes and secure data exchanges, leading to its adoption by major platforms and browsers \cite{rfc9000,zirngibl2024quic}. For example, 75\% of Facebook traffic was QUIC-based in 2020 after their successful migration of its apps, including Instagram, and server infrastructure to QUIC~\cite{facebook}. 

QUIC has multiple unique features such as 0-RTT handshaking, connection migration, and multiplexing~\cite{rfc9000}. QUIC’s innovative design significantly reduces the number of round trips needed to establish a connection \cite{rfc9000, rfc9001}. 
Its handshaking mechanism allows servers to respond to client requests without maintaining state between packets, thus improving performance \cite{rfc9000, rfc9002}. 
However, previous work has discovered that QUIC protocol is vulnerable to handshake flooding attacks~\cite{ teyssier2023empirical}, and the bottleneck of the QUIC server is most likely the CPU consumption under such attacks.
Attackers can exploit QUIC’s handshake mechanism by sending a large number of invalid or incomplete handshake requests. The server attempts to process those requests to establish connections~\cite{teyssier2023empirical,teyssier2023quicshield}. This consumes the server's resources quickly due to the high CPU amplification factor~\cite{ teyssier2023empirical}, leading to resource exhaustion with less effort from the attacker~\cite{nawrocki_quicsand:_2021, teyssier2023empirical}. 

In recent years, some research has been done to detect QUIC handshake flooding attacks effectively~\cite{teyssier2023quicshield, joarder2024quicwand, joarder2024quicpro}. However, existing mitigation techniques focus primarily on reactive measures, such as the monitoring and the filtering of malicious traffic. They do not fix the vulnerability from the root cause - high resource amplification factors during the attack. Instead, they may demand extra resource for the detection. furthermore, false positives and false negatives are unavoidable in attack detection, and they may compromise the performance of the network. As QUIC being increasingly adopted world-wide, addressing these handshake flooding attacks vulnerabilities with more proactive, preventive and cost-effective non-AI measures is imminent.

In this paper, we propose QFAM ({Q}UIC {F}looding {A}ttack 
Mitigation), a solution to address the DDoS vulnerability from the root cause by reducing the CPU amplification factor when the server is under attack. In QFAM, the QUIC server presents a cryptographic challenge to the QUIC client. The challenge can be integrated into its Retry packet during the handshaking process and it is only activated if the server switches to handshake flooding mitigation mode. The client must solve the cryptographic challenge and send the answer back to the server to finish the handshaking. The server takes a small effort to verify the solution from the client and will only move forward with the rest of the handshaking process if the answer from the client is correct. The server can tune the difficulty level of the challenge to balance the resource consumption between the server and the attacker. Our evaluation results on real testbed demonstrate that the attackers' cost of launching handshaking flooding is significantly increased after the server switches to the mitigation mode, therefore effectively mitigating the impact of 
the handshake flooding attack.  

The contributions of this paper can be summarized as follows:
\begin{itemize}
    \item We propose QFAM, a novel solution to mitigate QUIC  handshake flooding attacks by sending cryptographic challenges to the clients to reduce the CPU amplification factor during the attack.  
    \item We provide a detailed design on how to modify the RETRY TOKEN to effectively integrate a cryptographic challenge into the QUIC handshake protocol. 
    \item We evaluate our solution using our testbed with one QUIC server and multiple QUIC clients. Our experimental results demonstrate the effectiveness of our proposed solution in mitigating handshake flooding attacks with a low overhead for legitimate clients.
\end{itemize}

The rest of this paper is organized as follows. \textcolor{blue}{Section \ref{Related_Works}} provides a thorough review of existing strategies for detecting and mitigating handshake flooding attacks against QUIC. In \textcolor{blue}{Section \ref{Preliminaries}}, we present fundamental concepts of QUIC’s handshake mechanism and address validation mechanism. \textcolor{blue}{Section \ref{Mitigation_Scheme}} presents our proposed solution to mitigate QUIC handshake flooding attacks. \textcolor{blue}{Section \ref{Experiments_and_Evaluation}} describes the evaluation results followed by a discussion of our findings in \textcolor{blue}{Section \ref{Discussion}}. Lastly, \textcolor{blue}{Section \ref{Conclusion}} concludes the paper.

\section{Related Works}
\label{Related_Works}

QUIC \cite{thomson_version-independent_2021,rfc9000,rfc9002,duke_quic_2023} is invented particularly to reduce latency and enhance security by embedding TLS within the transport layer, enabling faster connections than TCP \cite{kakhki2019taking,chatzoglou2023revisiting,sengupta_accelerating_2024}. 
Its special handshake design remarkability reduces the number of round trips (RTTs) needed to establish a connection \cite{mucke2024reacked, nawrocki_interplay_2022}. This handshake technique
not only enhances latency and throughput but also makes the
QUIC protocol suitable for real-time applications and
services\cite{palmer2018quic,fernandez2024exploiting,de2017multipath}. However, this unique handshake mechanism design also makes it vulnerable to handshake DDoS flooding attacks \cite{sengupta_accelerating_2024,teyssier2023empirical}. These attacks often manipulate QUIC's address validation mechanism and version negotiation, permitting attackers to frequently initiate cryptographic procedures on the server to increase the CPU and memory load \cite{teyssier2023quicshield, teyssier2023empirical}. As the first initiative regarding handshake flooding, Teyssier et al. have shown that QUIC servers can experience a CPU amplification factor of up to 4.6x compared to TCP (TCP SYN Cookies), indicating that for every unit of CPU resources spent by an attacker, especially raising their vulnerability to resource exhaustion during such attacks \cite{teyssier2023empirical}.

Although QUIC includes defense mechanisms like address validation to mitigate resource exhaustion via DoS attacks, research by Nawrocki et al. \cite{nawrocki_quicsand:_2021} found that this mechanism is often disabled in practice, leaving many QUIC implementations. As a result, they can be attacked by QUIC handshake flooding \cite{teyssier2023empirical}. The authors also emphasized that while QUIC’s design restricts amplification attacks, the protocol remains vulnerable to resource exhaustion, mainly in real-world deployments where unsolicited Internet traffic often contains handshake DDoS flooding attempts \cite{nawrocki_quicsand:_2021}.

Multiple defense methods are published to address these handshake flooding attack-related vulnerabilities \cite{teyssier2023quicshield,joarder2024quicwand,joarder2024quicpro}. One of the solutions is "QUICShield". It employs Bloom Filters and Generalized Likelihood Ratio (GLR-CUSUM) algorithm to detect anomalies rapidly in handshake requests \cite{teyssier2023quicshield}. While QUICShield identifies malicious traffic patterns and incomplete handshakes, its static nature leads to a higher rate of false positives, reducing its adaptability in dynamic environments like real public networks. Additionally, QUICShield has limited capability of detection \cite{teyssier2023quicshield}. To overcome these shortcomings, "QUICwand" introduced a machine learning-based approach using Bayesian Optimization to dynamically adjust Bloom Filter parameters against QUIC handshake flooding attacks \cite{joarder2024quicwand}. This incredibly reduced false positive rates and enhanced detection accuracy by adapting to evolving traffic patterns and attack behaviours. However, both QUICShield and QUICwand leave servers vulnerable to initial resource exhaustion during handshake flooding attacks \cite{teyssier2023quicshield, joarder2024quicwand}.

To overcome the limitations of QUICShield and QUICwand, "QUICPro" introduced Deep Reinforcement Learning (DRL) to optimize defense techniques dynamically against handshake flooding attacks \cite{joarder2024quicpro}. Using DRL algorithms, such as Proximal Policy Optimization (PPO), QUICPro can learn from network behaviour and proactively adjust its defenses in real-time public networks\cite{joarder2024quicpro}. This allows the method to continuously monitor and adapt to changing traffic conditions, offering improved protection while reducing false positive rates \cite{joarder2024quicpro}. However, although QUICPro is costly in computation, it can not fully prevent initial resource exhaustion during handshake flooding attacks \cite{joarder2024quicpro}.

\section{Preliminaries}
\label{Preliminaries}

In this section, we provide an overview of the QUIC connection establishment mechanism, outlining both its fundamental process and potential vulnerabilities to flooding attacks. \textcolor{blue}{Table \ref{tab:acronyms}} summarizes the symbols and notation used throughout this section and the entire paper for clarity and consistency.



\begin{table}[ht]
\centering
\caption{Acronyms with Their Full Forms}
\begin{tabular}{|l|l|}
\hline
\textbf{Full Forms} & \textbf{Acronyms} \\ \hline
Fixed Bit & FB \\ \hline
Long Packet Type & LPT \\ \hline
Destination Connection ID Length & DCIDL \\ \hline
Destination Connection ID & DCID \\ \hline
Source Connection ID Length & SCIDL \\ \hline
Source Connection ID & SCID \\ \hline
Token Type & TT \\ \hline
Token Sequence & TS \\ \hline
Unique Token Number & UTN \\ \hline
Token Header & TH \\ \hline
Token Body & TB \\ \hline
Encrypted Token Body & ETB \\ \hline
Token Integrity Check Value & ICV \\ \hline
Encryption and Authentication Key & K \\ \hline
Initialization Vector & IV \\ \hline
Original Destination Connection ID Length & ODCIDL \\ \hline
Original Destination Connection ID & ODCID \\ \hline
Opaque Data & OD \\ \hline
Client IP Address & CIPA \\ \hline
Retry Source Connection ID Length & RSCIDL \\ \hline
Retry Source Connection ID & RSCID \\ \hline
Retry Token & RT \\ \hline
Token Identifier Number & TIN \\ \hline
Matched Random Number & MRN \\ \hline
Challenge Complexity Index & CCI \\ \hline
Difficulty Value & DV \\ \hline
Packet Number & PN \\ \hline
Associated Data & A \\ \hline
\end{tabular}
\label{tab:acronyms}
\end{table}

\subsection{Handshake Mechanism of QUIC}

QUIC handshake mechanism plays a key role in establishing secure communication between clients and servers while maintaining low-latency performance \cite{rfc9000, thomson_version-independent_2021}. There are two types of handshake mechanisms in QUIC: 1-RTT handshake and 0-RTT handshake, each offering different security and performance trade-offs \cite{rfc9000}. These two types of handshakes involve exchanges of cryptographic messages to negotiate encryption keys and ensure mutual authentication before any application data is transmitted. Below, we describe the handshake mechanisms and compare the security measures involved.

\begin{figure*}[htbp]
  \centering
  \includegraphics[width=\textwidth]{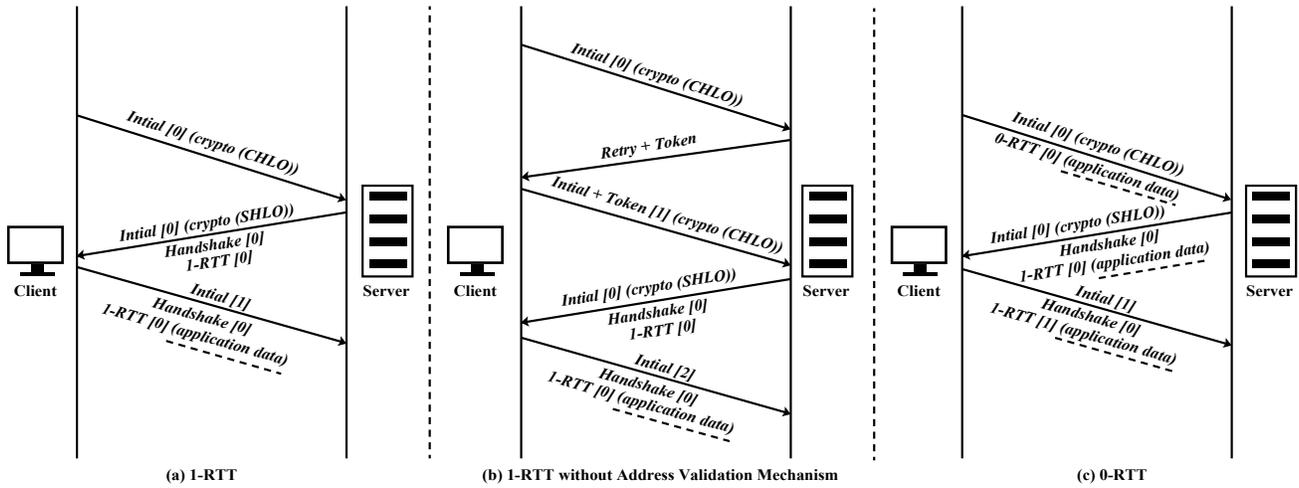} 
  \caption{1-RTT and 0-RTT Handshake Mechanisms in details; (a) 1-RTT handshake without address validation mechanism; (b) 1-RTT with address validation mechanism; (c) 0-RTT facilitating early data transmission}
  \Description{Diagram showing 1-RTT and 0-RTT handshake mechanisms with details of address validation and early data transmission.}
  \label{fig:main_handshake_mechanism}
\end{figure*}

In \textcolor{blue}{Figure~\ref{fig:main_handshake_mechanism}}, we observe three (3) different QUIC handshake scenarios, each offering varying levels of security and latency. The first handshake scenario, depicted in \textcolor{blue}{Figure~\ref{fig:main_handshake_mechanism}(a)}, represents the 1-RTT handshake without an Address Validation Mechanism (AVM). In this handshake process, the client and server exchange handshake messages before transmitting any application data. This handshake ensures that the cryptographic keys and credentials are appropriately negotiated, thus providing high security. The second handshake scenario, illustrated in \textcolor{blue}{Figure~\ref{fig:main_handshake_mechanism}(b)}, improves the 1-RTT handshake with an additional Address Validation Mechanism. In this mechanism, the server sends a Retry message (packet) with a token to the client. The client then uses this token in a subsequent handshake attempt to prove that it controls the claimed legitimate IP address. This additional step is particularly important for mitigating Denial-of-Service (DoS) attacks, where attackers might attempt to overwhelm the server with illegitimate connection requests. Finally, \textcolor{blue}{Figure~\ref{fig:main_handshake_mechanism}(c)} represents the 0-RTT handshake. Unlike the 1-RTT handshake, this mechanism allows the client to send application data immediately without waiting for the completion of the handshake. This reduces latency significantly, making it beneficial for session resumption. However, the 0-RTT handshake comes with security trade-offs, as early data transmission raises the risk of replay attacks, where an attacker might attempt to replay previous session data\cite{rfc9000}.

\subsection{Basics of QUIC Address Validation Mechanism}

The Address Validation Mechanism in QUIC is an important feature designed to prevent IP spoofing and mitigate certain types of flooding DDoS attacks \cite{rfc9000}. QUIC ensures that the client is reachable and minimizes the risk of processing illegitimate requests by validating the client’s IP address (see in \textcolor{blue}{Figure~\ref{fig:main_handshake_mechanism}(b)}). This mechanism protects servers from all types of DoS attacks that exploit the initial handshake process \cite{rfc9000}. When a client first attempts to connect, it can not yet have a validation token. The server responds to the initial ClientHello (CHLO) message with a Retry packet, including a token. The client needs to return this token with CHLO to the server to prove that it can receive messages at the claimed legitimate IP address \cite{rfc9000}. The client’s response with the token ensures its reachability at the provided valid IP address. The server uses this token to validate the IP address without storing it, making it a stateless mechanism. This special QUIC design permits the server to process multiple connection attempts proficiently without extreme memory usage\cite{rfc9000}.

\subsection{Attack Strategies of QUIC Handshake Flooding Attacks}

Handshake flooding attacks in QUIC exploit the protocol's reliance on cryptographic operations during the handshake phase, making it vulnerable to various DDoS attacks. Attackers target this initial stage of communication to exhaust server resources before completing a secure connection. Below, we will explore several ways attackers attempt to overwhelm or manipulate the QUIC handshake process.

\subsubsection{QUIC Handshake Flooding Attack in Absence of Address Validation}
\begin{figure}[htbp]
  \centering
  \includegraphics[width=0.9\linewidth]{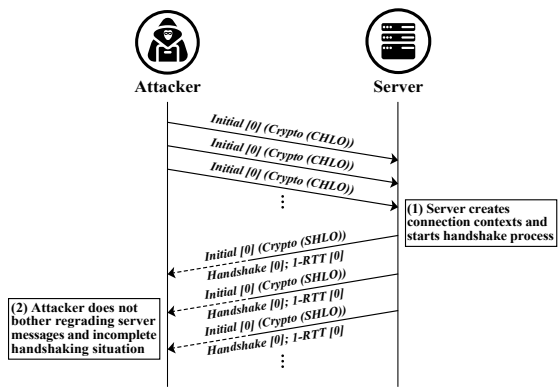}   
  \caption{Handshake flooding attack without QUIC Address Validation}
  \Description{}
  \label{fig:attack_dynamics}
\end{figure}

Since enabling the address validation mechanism adds one more communication round during the handshake process and connection establishment, it is often disabled on servers. Since the handshake process is an asymmetric process regarding computational overhead in the initial steps, the absence of this mechanism allows an attacker to initiate a handshake flooding attack even with a spoofed IP address \cite{nawrocki_quicsand:_2021}. Attackers can exploit this vulnerability by sending numerous CHLO messages, as depicted in \textcolor{blue}{Figure \ref{fig:attack_dynamics}}. 

Typically, upon receiving an Initial packet containing a CHLO message, the QUIC server engages in cryptographic operations to establish a secure connection, all of which impose a substantial computational load on the server \cite{rfc9000,rfc9001}.
The asymmetry of effort in this type of attack is significant: while the attacker only needs to send small CHLO packets, the server must perform intensive cryptographic tasks for each handshake. As a result, the server’s CPU and memory resources are rapidly consumed, leading to performance degradation and, eventually, a denial of service.

As illustrated in \textcolor{blue}{Figure \ref{fig:attack_dynamics}}, the continuous flood of CHLO messages overwhelms the server, forcing it to allocate resources repeatedly for connections that are never fully complete. This attack vector is particularly effective against QUIC servers that lack robust address validation mechanisms. By exhausting server resources, attackers can effectively disrupt service availability, thereby degrading the performance of legitimate client connections.

\subsubsection{QUIC Handshake Flooding Attack with Address Validation Mechanism}
\begin{figure}[htbp]
  \centering
  \includegraphics[width=0.9\linewidth]{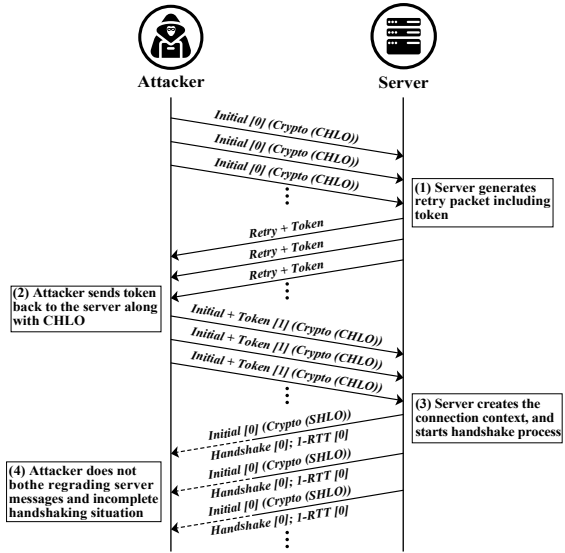}
  \caption{Handshake Flooding Attack With QUIC Address Validation}  
  \Description{}
  \label{fig:handshake_flooding_with_validation}
\end{figure}

The QUIC handshake integrates the negotiation of cryptographic and transport parameters. Address validation ensures that an endpoint cannot be used for traffic amplification attacks, where a packet is sent to a server with a spoofed source address that identifies a victim \cite{rfc9000}. However, even with QUIC address validation enabled, an attacker may still initiate a handshake flooding attack if they control the spoofed IP address.
In this scenario, when a client initiates a connection, it sends an Initial packet containing a CHLO message. Upon receiving the CHLO, the server responds with a Retry packet that includes an address validation token. This token is cryptographically tied to the client’s IP address to prevent spoofing. Consequently, the server generates Retry packets with these tokens and sends them back to the attacker. The attacker can then respond with a new Initial packet that contains both the CHLO and the server-issued token.
The server then proceeds with the next steps of the handshake, including the Server Hello (SHLO), which requires resource-intensive cryptographic operations and key generation for application data encryption. In this scenario, the attacker can still execute a handshake flooding attack similarly to before, with the exception that they must perform one additional communication round.
As depicted in \textcolor{blue}{Figure \ref{fig:handshake_flooding_with_validation}}, attackers can continuously send Initial packets with replayed tokens, forcing the server to repeatedly validate tokens and generate handshake keys. This process consumes significant server resources while the connections remain incomplete, causing a substantial increase in server load. The result is a high amplification factor, where minimal effort from the client side leads to significant computational demands on the server, effectively degrading its performance and potentially leading to denial of service.

\subsubsection{Amplifying the CPU Amplification Effect Through Computationally Intensive Cryptographic Algorithms}
Attack Amplification factors(AFF) are metrics to measure the level of leverage attackers can achieve when launching DDoS~\cite{rossow2014amplification}. It is measured by the resource the server dedicates to respond to the requests from the attacker, compared the resource attacker uses to generate the attack requests. For example, the amplification factor for Smurf attack can be as high as $255$~\cite{kumar2007smurf}. Although AFF typically refers to the bandwidth amplification, CPU/memory amplifications are also important since more sophisticated DDoS attacks target CPU-exhaustion~\cite{meng2018rampart,teyssier2023empirical} or memory-exhaustion~\cite{du2024medusa,pietrantuono2023survivability}. Previous study measured a CPU amplification factor of $4.6$ under handshake flooding attacks under the default settings on \textit{aioquic}~\cite{teyssier2023empirical}.

Since QUIC and TLS 1.3 support various cryptographic algorithms for key establishment, an attacker can exploit this flexibility to further enhance the CPU amplification of handshake flooding attacks. By selecting algorithms with high computational overhead, attackers can intensify the attack's strength, forcing the server to expend significant CPU resources. Attackers can further amplify this effect by sending pre-calculated or random values as their key shares during connection establishment, instead of generating legitimate values as required by the cryptographic process. This approach increases the computational burden on the server, effectively maximizing the CPU amplification impact of the attack. In this regard, we adopt and incorporate this approach throughout this study.

\begin{figure}[H]
  \centering \includegraphics[width=0.7\linewidth]{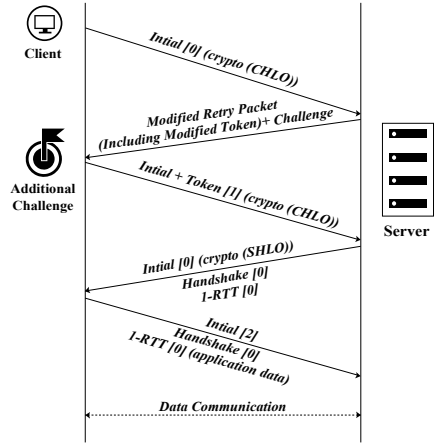} 
  \caption{Illustration of QFAM}
    \Description{}
  \label{fig:quicprotector}
\end{figure}

\section{Proposed Mitigation Scheme}
\label{Mitigation_Scheme}

In this section, we present QFAM, our proposed mitigation approach against QUIC handshake flooding attacks. Different from the conventional reactive approach to detect and block attack traffic~\cite{mishra2023containing,pietrantuono2023survivability,teyssier2023quicshield}, this approach mitigates attacks by increasing the cost of attackers. Therefore, our approach does not suffer from the consequence of false positives or false negatives, and it can used in combination with the conventional approach if desired. The idea is to poses a cryptographic challenge to the client to increase the CPU cost.
The cryptographic challenge can be delivered to the client through the token in the RETRY packet. 
This ensures that only legitimate clients—those who are reachable and capable of solving the cryptographic challenge—receive high priority on the server side during the handshake process. \textcolor{blue}{Figure \ref{fig:quicprotector}} provides an overview of the QFAM process. By employing this scheme, the server can mitigate the effects of CPU amplification and prevent overload caused by flooding attacks. To achieve these objectives, we enhanced the RETRY packet structure and strengthened the address validation mechanism. The detailed methodology follows in the steps below:

\subsection{Enhanced RETRY Packet Structure with Cryptographic Challenge}

QUIC protocol's standard RETRY packet, as defined in RFC 9000 \cite{rfc9000}, is structured to send a secrete to the client to verify the IP address. Our enhanced RETRY packet incorporates a cryptographic challenge into the RETRY packet so that it serves both the purpose of IP verification and resource balancing. Below, we compare the original RFC 9000 RETRY packet structure with our enhanced version.

\subsubsection{Original RFC 9000 RETRY packet structure} 
\begin{figure*}[ht]
\centering
\includegraphics[width=\textwidth]{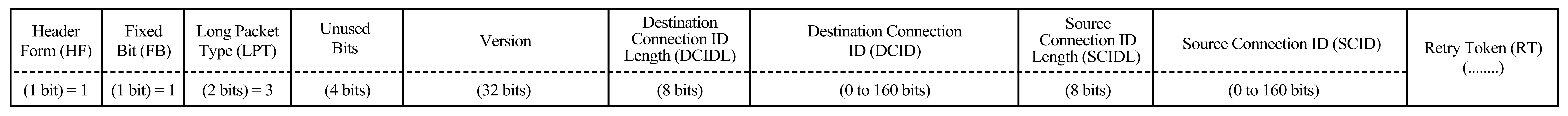}   
\caption{Structure of the QUIC RETRY Packet}
  \Description{}
\label{fig:before_retry_packet_structure}
\end{figure*}

The RETRY packet structure, as defined in RFC 9000, is illustrated in \textcolor{blue}{Figure ~\ref{fig:before_retry_packet_structure}}. This packet is generated by the server and includes several fields, notably an unused field 'Unused Bits' and the 'Retry Token'. The 'Unused Bits' field is 4 bits long and reserved for potential future extensions or protocol modifications. The Retry Token is created by the server and sent to the client, who must return this token to the server immediately. Upon receiving the token, the server verify the validity of the token and decide whether to move forward with the rest of the handshake process. Similar address validation mechanism is used in TCP handshaking.


\subsubsection{Enhanced RETRY Packet Structure} \label{sec:enhanced}
\begin{figure*}[ht]
\centering
\includegraphics[width=\textwidth]{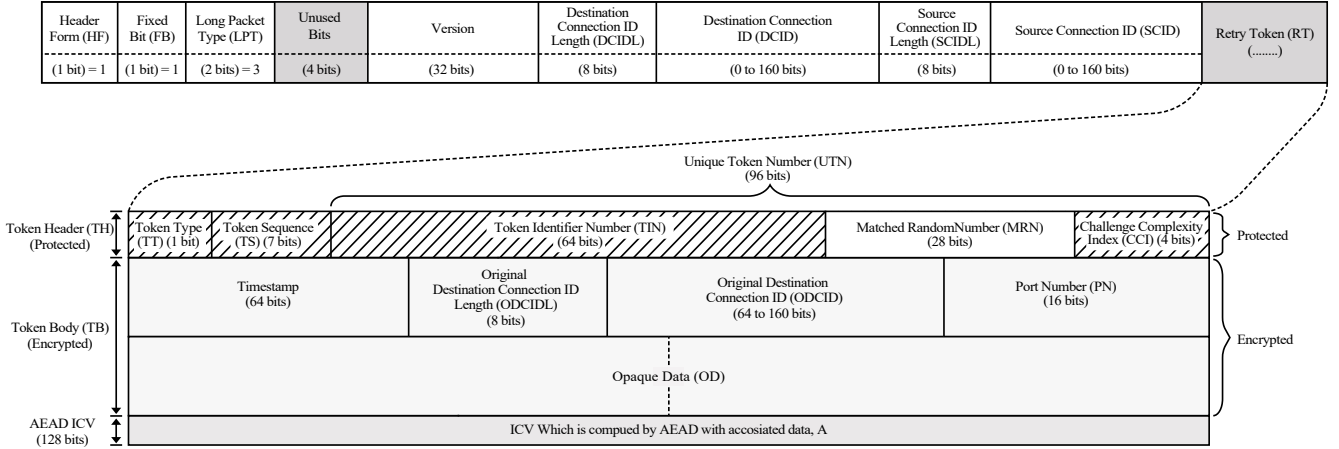}   
\caption{Enhanced structure of the RETRY packet and its corresponding Retry Token in QFAM}
  \Description{}
\label{fig:retry_packet_structure}
\end{figure*}

In order to ensure compatibility with the original version of RETRY packets, we make a slight modification. This modification mainly affects the Retry Token and unused field in the RETRY packet structure. The structure of the enhanced packet is demonstrated in \textcolor{blue}{Figure ~\ref{fig:retry_packet_structure}}.

To differentiate the enhanced version of the RETRY packet from the original, we set a bit in the unused field, such as the most significant bit. We name the bit \textit{mitigation bit}. During connection establishment, the server sets the mitigation bit to indicate the packet as an enhanced RETRY packet. When a client receives a RETRY packet with the mitigation bit on, it should process the packet according to the proposed procedure as described in Section~\ref{sec:client-handling-retry-packet}.


\subsubsection{Key Modifications and Enhanced Retry Token} \label{Token_Specification}

We modified the RETRY token to mitigate flooding attacks. Our proposed RETRY token structure is shown in \textcolor{blue}{Figure \ref{fig:retry_packet_structure}}, and it is compatible with { \cite{I-D.ietf-quic-retry-offload}. The details are as follows:


A Retry token consists of three parts: the Token Header, Token Body (TB), and Token Integrity Check Value (ICV). The Token Header includes a 1-bit token type identifier, a 7-bit Token Sequence, and a 96-bit Unique Token Number.
The token type identifier signifies the token type, indicating a Retry token if set to '0', or a NEW-TOKEN otherwise. The 7-bit Token Key Sequence Identifier is used by the server to retrieve the token encryption keys and IVs. 

To implement our proposed approach without significant changes to the protocol or token structure, we divide the Unique Token Number (96 bits) into three parts:
\begin{itemize}
    \item 64 bits are reserved for protocol-specific use and are generated by the server, serving as the Token Identifier Number (\(TIN\)). This value is set to a random number generated using a cryptographically secure random number generator.
    \item 28 bits are allocated for a Matched Random Number (\(MRN\)), which serves as the solution to the cryptographic challenge.
    \item 4 bits are allocated for the Challenge Complexity Index (\(CCI\)), determining the complexity of the cryptographic challenge.
\end{itemize}
The server uses this 4-bit challenge complexity index to dynamically adjust the difficulty of the cryptographic challenge presented to clients. For example, the complexity can be represented as a binary number ranging from '0000' (the lowest difficulty level) to '1111' (the highest level, corresponding to decimal 15). When the server is under low stress and the number of connection establishment requests is low, indicating a low risk of attack, it may set a lower difficulty level. However, during handshake flooding attack
the server may increases the challenge complexity level, imposing a higher computational overhead on attackers. 
The Token Header is transmitted in clear text but protected using AES128-GCM, as part of the AEAD associated data.

The Token Body consists of a 64-bit Timestamp, in network byte order, which indicates the expiration time of the Retry token; an 8-bit Original Destination Connection ID Length (ODCIL); the Original Destination Connection ID (ODCID), copied from the corresponding field in the client's Initial packet; a 16-bit Source Port number, representing the UDP datagram port that triggered the Retry packet; and Opaque Data, which may be used by the server to encode additional optional information. The entire Token Body is encrypted using AES128-GCM.

The Token Integrity Check Value is 128 bits in length and is computed to protect the token's integrity. It is derived from the Client IP Address (128 bits), Token Type (1 bit), Token Key Sequence (7 bits), Token Identifier Number(TIN)(64 bits),
Challenge Complexity Index (CCI) (4 bits), Retry Source Connection ID Length (RSCIDL) (8 bits), and the Retry Source Connection ID (up to 160 bits). These values are incorporated as the associated data, $A$, in the AEAD encryption process.
\[
A =  \left( 
\text{CIPA} \, (128) \, || \,
\text{TT} \, (1) \, || \,
\text{TS} \, (7) \, || \,
\text{TIN} \, (64) \, || \right.
\]
\[
\left. 
\text{CCI} \, (4) \, || \,
\text{RSCIDL} \, (8) \, || \,
\text{RSCID} \, (160)
\right)
\]
The token is encrypted as follows:
\begin{equation}
\begin{aligned}
(ETB, ICV) = AEDA_{Encrypt}(K, IV, A, TB)
\end{aligned}
\label{eq:data_composition1}
\end{equation}

Where $ETB$ represents the encrypted token body, $ICV$ is the integrity check value, $K$ is the encryption and authentication key, $IV$ is the initialization vector, $A$ is the associated data, and $TB$ is the token body.

In the following, we outline the necessary steps the server must take to implement the proposed mitigation technique, including the generation and management of the enhanced RETRY tokens. Additionally, we detail how the client should handle these tokens in response.






\subsection{Server's Process for Generating and Sending Enhanced RETRY Packet}
\label{sec:sending-enhanced-retry-packet}

Whenever the server perceives a handshake flooding attack, based on dynamic analysis of traffic patterns and threat assessment, it may initiate the mitigation mechanism by generating and sending the proposed enhanced RETRY packet. This process is triggered by setting a bit in the unused field of the RETRY packet to indicate that this packet is an enhanced RETRY packet. Depending on the server load and real-time traffic analysis, the server generates a RETRY token, as demonstrated in subsection \ref{Token_Specification}, and sets the challenge complexity value in the defined field (CCI). Upon crafting the RETRY packet with these parameters, the server transmits it to the client, who must then resolve the cryptographic challenge to proceed with the secure handshake.





\subsection{Client Handling of Enhanced RETRY Packet in QFAM}
\label{sec:client-handling-retry-packet}

Upon receiving a RETRY packet, the client checks the unused field to determine whether it is an enhanced RETRY packet. If so, it extracts the challenge complexity value, the TIN, and the ICV from the received token. It then proceeds with the following steps.

\begin{enumerate}
    \item The client generates a new 28-bit random number $R$ and computes the value $Z$ as follows:
    \begin{equation}
\begin{aligned}
\text{Z} = h (& \text{ICV} \parallel \text{Token Identifier Number} \parallel \text{Port Number} \\
             & \parallel \text{Challenge Complexity Index} \parallel R)
\end{aligned}
\label{eq:data_composition2}
\end{equation}
    \item The client examines the challenge complexity value to determine the required number of leading zeros in the computed value \(Z\). If the computed 
$Z$ does not meet this requirement, the client iteratively recalculates $Z$ with different values of 
$R$ until it finds a value that produces the required number of leading zeros.
    \item Upon successfully finding such an $R$, the client inserts this value into the Matched Random Number field of the extracted RETRY token and sends this token back to the server in the ClientHello message, as specified in the QUIC protocol.
\end{enumerate}
 These steps also demonstrated in the Algorithm \ref{alg: A1}.

\subsection{Server Validation of Enhanced RETRY Token}
\label{sec:Server_Validation_of_Enhanced_RETRY_Token}

Upon receiving a new ClientHello message along with an enhanced RETRY token, the server extracts the Token key Sequence and, based on its value, retrieves the secret keys and IV used for encrypting the original token. It then extracts the Token Type, Token Identifier Number, and challenge complexity value from the received token and uses them as associated data, along with the Client IP Address, Retry Source Connection ID Length (RSCIDL), and the Retry Source Connection ID from the received ClientHello request, to validate the token.

If the token is valid and corresponds to the received ClientHello packet, the server checks whether the solution to the challenge is correct. To do this, it extracts the ICV, Token Identifier Number (64 bits), challenge complexity value, and the solution \(MRN\) (28 bits), as well as the client’s port number that initiated this ClientHello packet, and computes:

\begin{equation}
\begin{aligned}
Z' = h (& \text{ICV} \parallel \text{Token Identifier Number} \parallel \text{Port Number} \\
             & \parallel \text{challenge complexity value} \parallel MRN)
\end{aligned}
\label{eq:data_composition3}
\end{equation}

The server then verifies whether \( Z' \) contains the required number of leading zeros, as specified by the Challenge Complexity Index field in the token. If so, the server proceeds to establish a secure connection based on the QUIC protocol; otherwise, it rejects the request.

}


\begin{algorithm}[t]
\SetAlgoLined
\DontPrintSemicolon
\SetInd{0.7em}{0.9em} 
\small
\KwIn{(Retry Token)}
\KwOut{Matched Random Number $(MRN)$ \;
\rule{0.8\linewidth}{0.2pt}} 
$(ICV, TIN, PN, CCI) \xleftarrow{\text{Extracts}} \text{Retry Token}$ \;
$R \longleftarrow 0$\;
\While{true}{
$Z \longleftarrow h(ICV\parallel TIN \parallel PN \parallel CCI \parallel R)$\;
{\If{$Z$ has $|CCI|$ or more leading zeros}{
     \textbf{break}}
}
$R \longleftarrow R+1$\;
}
$MRN \longleftarrow R$ \;
\Return $MRN$ 

\caption{Client-Side Cryptographic Challenge Solution} \label{alg: A1}
\end{algorithm}

\subsection{Complexity Analysis} \label{sec:complexity}
The purpose of the aforementioned cryptographic challenge was two fold: 1) to verify the client has the IP address it claimed in the initial CHLO; 2) to increase the cost on the client for handshaking. Due to the nature of the challenge, the expected time to compute the result should increase exponentially with the difficulty level $n$. That means the computation complexity of Algorithm 1 is $O(2^n)$ for the client to solve the challenge. On the other hand, the server uses constant time to generate the challenge and verify the answer returned from the client. The computation complexity on the server side is $O(1)$. As a result, when the difficulty level $n$ increases, the client needs to invest more CPU time than the server during the handshake. When an appropriate difficulty level $n$ is chosen, the CPU resource consumption between the attacker and the server can be balanced. This way, the attacker's power can be significantly reduced, resulting in the mitigation of the handshake flooding attacks. 

\subsection{The Mitigation Mode}
Its worth noting that the cryptography challenges are only used when the server detects handshake flooding attacks. When there is no attack, no one is required to solve any challenge. This way, the performance of the normal clients would not be impacted when there is no handshake flooding attack.  When a handshake flooding attack is detected, the server turns on the mitigation mode by setting the mitigation bit on (Section~\ref{sec:enhanced}). 

When the mitigation mode is on, the server requests each client to solve a cryptographic challenge, and only proceeds with the key share computation and sends SHLO packet when the client proves that it solves the challenge correctly. If the client does not solve the challenge correctly, the server could either terminates the handshake process immediately, or sends the CHLO to a low priority queue to be processed when the resource allows. The later decision can be made to take care of normal clients which runs on a version that does not support mitigation mode.

\section{Experiments and Evaluation}
\label{Experiments_and_Evaluation}

In order to evaluate the effectiveness of the proposed scheme, we deployed our mitigation mechanism on our testbed with QUIC clients and a QUIC server running \textit{aioquic}~\cite{aioquic2023}. \textit{Aioquic} is a Python library for the QUIC network protocol, featuring a minimal TLS 1.3 implementation, a QUIC stack, and an HTTP/3 stack. We customized \textit{aioquic} to enable the server to generate and manage enhanced RETRY packets, while clients handle these packets accordingly. Various parameters were measured from both the clients and server processes to assess the impact of the proposed scheme on the QUIC protocol.

\subsection{Evaluation Environments}
\label{Evaluation_Environments}

We deployed our proposed mitigation technique on \textit{aioquic} within a private local network as a testbed with three machines, one QUIC server and two clients (a attacker and a normal client). The attacker client machine was configured to run a modified version of \textit{aioquic} to simulate a flooding attack scenario effectively.  All the clients and server processes were also modified to implement the DDoS mitigation mechanism. The specifications of the evaluation environment and the parameters used in this testbed are presented in Table~\ref{Table:2}.

For the hardware platforms, we kept one processor core active and disabled the remaining cores on the client, attacker, and server machines. Additionally, the Linux command \textit{ps} was used on all machines to monitor system performance and measure resource utilization.

\begin{table}
\caption{Platforms and Parameters Used in the Evaluation Environment}\label{Table:2}
\begin{tabular}{p{3.4cm}p{4.2cm}}
\toprule
 Parameter & Description \\
\midrule
OS for \textit{aioquic} Client and Server & Ubuntu 22.04 LTS \\ Hardware Platform for \textit{aioquic} Client and Server & Processor: 13th Gen Intel Core i7-13700K, RAM: 32 GB \\ Network Device & TP-Link Router \\
Applied Cryptography Algorithm for Attacker & SECP384R1 \\
\bottomrule
\end{tabular}
\end{table}

\subsection{Evaluation Results}
\label{Evaluation_Results}
In this subsection, we present the results of our experimental evaluation. We demonstrate the impact of the proposed mitigation scheme on the QUIC protocol across various scenarios. The details are as follows:

\subsubsection{Impact of Increasing Attack Rate on CPU Usage and CPU Amplification Factor} 
We designed an attack scenario where the attacker floods the server by sending numerous connection establishment requests, overloading the server's computational capacity. Since the QUIC protocol is vulnerable to CPU exhaustion and amplification attacks, the attacker can exploit this by sending customized ClientHello messages to overwhelm the server.

To intensify the attack, we modified the ClientHello messages sent by the attacker. Various cryptographic algorithms for key exchange and digital signatures, including X25519, X448, GREASE, SECP256R1, and SECP384R1, are supported in the \textit{aioquic} protocol. To further amplify the attack, the attacker was configured to use a precomputed public key share based on Group SECP384R1, which saves time further compared to computing the key share for every request, while the server has to complete the expensive key computation as part of the key exchange protocol. Given that the computational overhead for generating public key shares in Group SECP384R1 is higher than other groups, the attacker specifically chooses SECP384R1 for key exchange to maximize the CPU amplification.

\begin{figure}[tb]
\centerline{\includegraphics[width=0.50 \textwidth]{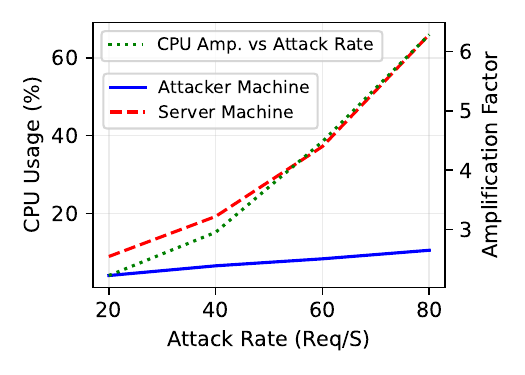}}
\caption{Impact of Increasing Attack Rate on CPU Usage and CPU Amplification Factor}
  \Description{}
\label{fig:F4}
\end{figure}

We ran the attacker and server on separate machines with identical characteristics, as shown in Table~\ref{Table:2}. The attacker sent connection establishment requests to the server, varying from 20 requests per second (Req/s) to 80 Req/s, as illustrated in Fig~\ref{fig:F4}. We observed that as the attack rate (number of connection establishment requests per second) increased, the CPU utilization on the attacker’s side rose slightly, while it increased significantly on the server’s side. We did not plot the results for 100 Req/s since the server was fully saturated and the measurement on the server could not be conducted effectively. Additionally, the results show the relationship between CPU amplification factor and attack rates. As we can see, the CPU amplification factor raised quickly as the attack rate increases and it reached over $6$ when the attack rate was 80 Req/s. We speculate the reason of this non-linear growing of CPU usage on the server side to be the extra resource required to maintain the increasing backlog of requests to be processed.

\subsubsection{Impact of Challenge Complexity on CPU Usage for Client and Server}
\begin{figure}[tbp]
  \centering
  \includegraphics[width=0.9\linewidth]{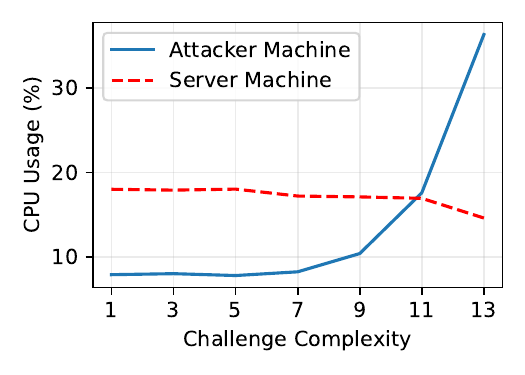} 
  \caption{Impact of Challenge Complexity on CPU Usage for Attacker and Server}
    \Description{}
  \label{Fig: Cpu_usage_AS}
\end{figure}

As explained in the last subsection, an attacker can carefully craft ClientHello packets using pre-computed key share to launch a handshake flooding attack on the server. To counter this, the server can activate the mitigation mode and send cryptographic challenges. To assess the impact of the challenge complexity (difficulty level), we set up an experimental environment consisting of one client and one server with the mitigation mechanism integrated. The client was configured to send requests at a constant rate of 40 requests per second. The server adjusts the challenge complexity to reduce the CPU amplification effect. The results of this experiment are illustrated in \textcolor{blue}{Figure \ref{Fig: Cpu_usage_AS}}. As shown, increasing the challenge complexity significantly raises the computational overhead on the attacker’s side. Therefore, when the server detects an attack, it can initiate the proposed mechanism, adjust the challenge complexity, and effectively mitigate the attack.

If the attacker intends to exploit the amplification factor and impose the burden of connection establishment's computational overhead on the server, they must first solve the challenge to proceed with the handshake process. By solving the challenge, the computational overhead on the attacker's machine increases, which eliminates the CPU advantage of the attacker during the handshake process. If the attacker chooses not to compute the cryptographic challenge, its CHLO would not be processed normally. Therefore, the high CPU amplification could not be achieved either.


\subsubsection{ Impact of Increasing Challenge Complexity on
Attack Rate Reduction}

\begin{figure}[tbp]
  \centering
  \includegraphics[width=0.9\linewidth]{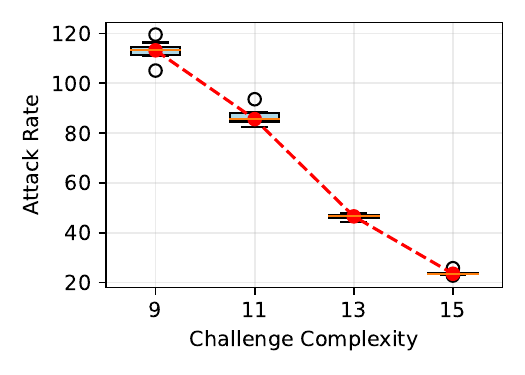} 
  \caption{Impact of Increasing Challenge Complexity on Attack Rate Reduction}
    \Description{}
    \label{fig:Attach_rate_time_with_median}
\end{figure}

In this scenario, we configured the attacker’s machine to send as many connection establishment requests as possible while simultaneously increasing the challenge complexity on the server side. The results of this experiment are illustrated in \textcolor{blue}{Figure \ref{fig:Attach_rate_time_with_median}}. As the figure shows, increasing the challenge complexity value leads to a corresponding decrease in the attack rate. This is because when the difficulty level of the challenges increases, the client dedicates more CPU time on solving the challenges for each connection request. Therefore, the number of connection requests it can support decreases. Therefore, the handshake flooding attack would be mitigated.

It is worth noting that this result is achieved under the assumption that the attacker chooses to cope with the challenges and sends the correct answers back to the server to proceed with handshaking. If the attacker chooses not to solve the challenges, then its attack rate does not change with the challenge
complexity. 
However, the server would not dedicate resources to process its CHLOs. Therefore, the handshake flooding attack would be mitigated.

\subsubsection{Evaluating CPU Usage Under Different Attack Rates and Mitigation Scenarios}

\begin{figure}[tbp]
\centerline{\includegraphics[width=0.50 \textwidth]{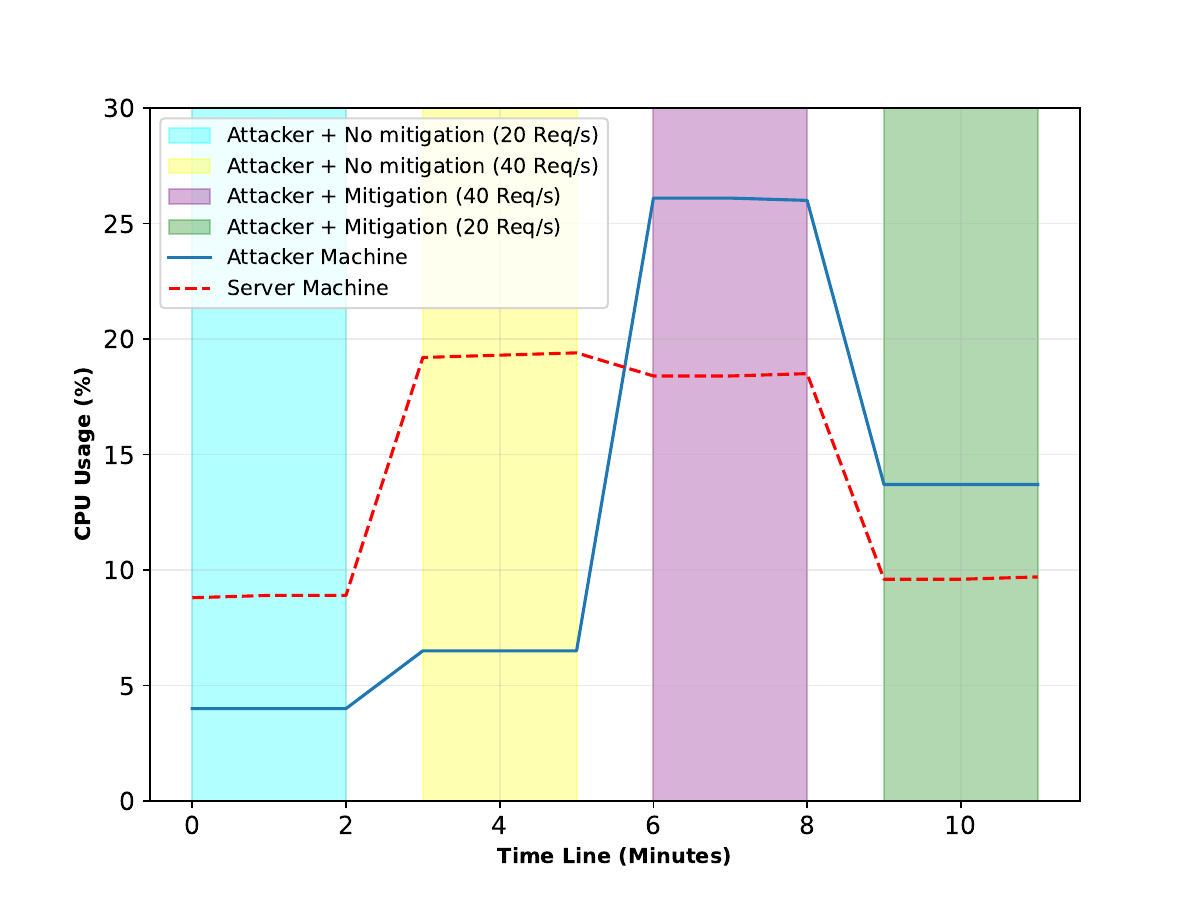}}
\caption{CPU Usage Comparison Between Attacker and Server in Non-Mitigation and Mitigation Modes (Challenge Complexity: $1100$)}
  \Description{}
\label{fig:Cpu_usage_Diff}
\end{figure}

As we discussed earlier, by carefully choosing the challenge complexity, the server can effectively mitigate handshake flooding attacks and counteract the CPU amplification effect exploited by attackers. To demonstrate the impact of the proposed mitigation scheme, we measured the CPU utilization on both the attacker's and server's machines across various scenarios, as illustrated in \textcolor{blue}{Figure \ref{fig:Cpu_usage_Diff}}.
In the first scenario, lasting for 2 minutes (from time 0 to 2 minutes), the mitigation technique was disabled, and the attacker continuously sent 20 connection-establishment requests per second to the server. During this period, the server's CPU overhead was more than twice that of the attacker's machine.
In the second scenario (minutes 3 to 5), the attack rate doubled to 40 Req/s. As expected, CPU usage on the server increased further compared to the observed rise in CPU usage on the attacker's machine.

From minute 6 to minute 11, we enabled the mitigation technique and set the Challenge Complexity Index to $1100$ (or 12 in decimal). As observed between minutes 6 and 8, this caused a significant increase in CPU usage on the attacker’s machine, while maintaining the attack rate at 40 Req/s. A small decrease in the server's CPU utilization was noted, due to the attacker's inability to solve the challenges associated with each address validation request sent by the server, thereby limiting the attacker's capacity to maintain the attack rate effectively.
In the final scenario, we reduced the attack rate back to 20 Req/s. As shown in the figure, the CPU usage on both the attacker’s and server’s machines decreased. However, the attacker’s CPU usage was higher compared to the earlier scenario with the same attack rate, reflecting the added computational burden of solving the challenges. Meanwhile, the server’s CPU usage returned to its previous level, with a slight increase attributed to the overhead of managing the enhanced retry packets.

\begin{figure}[tbp]
\centerline{\includegraphics[width=0.50 \textwidth]{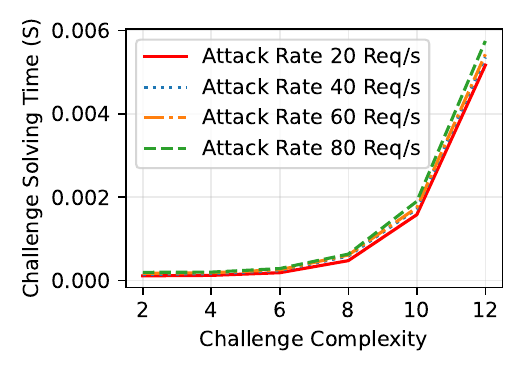}}
\caption{Impact of Challenge Complexity on Solving Time Across Different Attack Rates}
  \Description{}
\label{fig:Challenge_solving_time}
\end{figure}


\begin{figure}[tbp]
\centerline{\includegraphics[width=0.45 \textwidth]{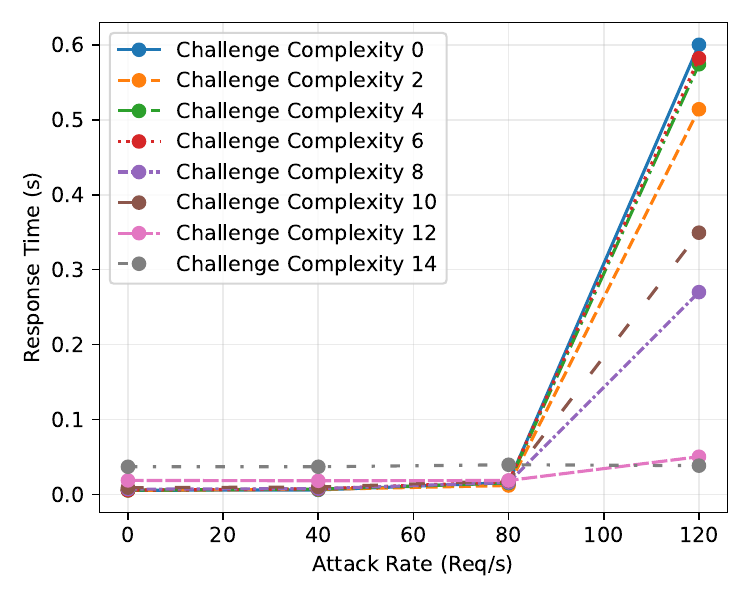}}
\caption{Effect of Varying Attack Rates on Response Time for Legitimate Clients}
  \Description{}
\label{fig:Response_time}
\end{figure}

\subsubsection{ Impact of Challenge Complexity on Solving Time}

Increasing the challenge complexity leads to a corresponding increase in both the challenge-solving time and the CPU usage on the attacker's side. In this experiment, we study the relationship between the challenge complexity and the challenge solving time. We set up an environment with one server and one attack client. The attacker sends handshake requests to the server server, and the server replies with cryptographic challenges. We measure the average challenge solving time on the attacker's side.
\textcolor{blue}{Figure \ref{fig:Challenge_solving_time}} illustrates the challenge-solving time across different attack rates and various challenge complexity levels. As observed, for a given challenge complexity value, the challenge-solving time remains relatively consistent across different attack rates. Furthermore, as the challenge complexity increases, the time required to solve the challenge also increases. This empirical results match our complexity analysis results in Section~\ref{sec:complexity}.

\subsubsection{Impact of Varying Attack Rates Across Different Challenge Complexity Levels on Response Time for Legitimate Clients}

To assess the impact of handshake flooding attacks on legitimate clients, we measured the response time for each request made by a legitimate client while an attacker launched varying attack rates against the server. 

The results are shown in \textcolor{blue}{Figure ~\ref{fig:Response_time}}. As illustrated, the response time for legitimate clients increases significantly in the presence of an attacker, particularly at higher attack rates and lower challenge complexity levels, such as at an attack rate of 120 requests per second (Req/s) with a challenge complexity of 0. At the same attack rate, as the challenge complexity level increases, the response time decreases. This decrease is especially evident at high attack rates (e.g., 120 Req/s) with a challenge complexity of 14. When the challenge complexity level is 14 since the server is not overloaded, the response time remains nearly constant. However, for lower attack rates with a challenge complexity of 14, the response time is slightly higher compared to other challenge complexity levels. This is because the legitimate client must solve a challenge of higher complexity, which takes more time than challenges of lower complexity.

\section{Discussion}
\label{Discussion}
In this study, we introduced a novel mitigation mechanism designed to counter QUIC handshake flooding attacks, especially effective in conditions where servers face high attack rates and substantial overload. 

This approach proposes modifying the RETRY token in the RETRY packet and integrating a cryptographic challenge into it. It serves as a complementary enhancement to the QUIC protocol's address validation mechanism.
By enabling this mechanism, the server can mitigate the impact of the CPU amplification effect by requiring the client to solve cryptographic challenges during the handshake process to gain high priority on the server side. If the client does not support this mitigation mechanism, it can still establish a connection with the server, but such requests will be handled with lower priority. In this case, the client treats the enhanced RETRY token as a standard token based on the original QUIC protocol and immediately sends it back to the server.

However, applying the proposed mechanism to lightweight or resource-constrained devices may present certain challenges. These devices may experience increased computational overhead, potentially leading to longer response times during the connection establishment phase due to the high processing demands associated with generating cryptographic key shares and solving the cryptographic challenge. Nonetheless, our LAN-based experiment demonstrates that the response times for connection establishment are substantially lower in scenarios with a high attack rate when the mitigation mechanism is applied compared to scenarios without any mitigation.

It is important to note that during high-rate flooding attacks, the server may drop or lose some connection establishment packets due to overload. In these situations, the client will typically retransmit its requests after the expiration of the Probe Timeout (PTO). In LAN environments, the PTO is generally shorter than in public networks, such as the Internet, allowing clients to reattempt connections more quickly. Consequently, the PTO in public networks can result in longer response times, especially under high attack rates, as clients experience delays in reconnection attempts.

In real-world applications without any mitigation strategies, response times would be significantly longer during high attack conditions. Our findings suggest that even though lightweight devices may require additional processing time to solve computational challenges, the overall delay perceived by users remains lower compared to scenarios without any mitigation measures. Further experiments are needed to confirm this under diverse real-world conditions.

Energy consumption is another concern, particularly for mobile devices. While solving computational challenges does increase energy use slightly, transmission operations often require more energy than computation \cite {10.1145/332833.332838}. In high-attack scenarios,  resending requests by clients may consume more energy than processing a single computational challenge. This observation highlights the need for further research to quantify energy demands under various attack intensities and to experimentally validate this hypothesis.

The limitation of our work is that we only conducted the experiments under \textit{aioquic}. As our future work, we intend to extend our experiment across different QUIC implementations such as picoquic~\cite{picoquic} and quic-go~\cite{quicgo}. Therefore, we can have a better understanding on the effectiveness of QFAM on QUIC systems implemented by different programming languages.

\section{Conclusion}
\label{Conclusion}

In this paper, we presented QFAM, a cryptographic defense mechanism to mitigate handshake flooding attacks in QUIC. We proposed a modification to the QUIC Retry packet structure and developed an enhanced Retry token by incorporating a cryptographic challenge. This enhancement is compatible with QUIC, increases computational demands on the attacker side, and protects server resources from exhaustion. Our experiments and evaluation results demonstrate that the proposed scheme is particularly effective in high handshake flooding attack scenarios, conserving server resources and reducing CPU amplification effects.

Future work will explore its performance in varied network conditions, including public networks, and assess its energy impact on mobile and IoT devices to optimize its applicability, ultimately enhancing the resilience and efficiency of QUIC deployments.


\appendix

\section{Ethics}
This work does not raise any ethical issues.


\end{document}